\documentclass[aps,prb,twocolumn,showpacs,floatfix]{revtex4}
\usepackage{graphicx}
\usepackage{xcolor}
\usepackage{ulem}

\def\fecr{FeCr$_{2}$S$_{4}$}
\def\Sc{FeSc$_{2}$S$_{4}$}

\begin{document}

\title{Comparative study of \fecr\ and \Sc: Spinels with orbitally active A site}
\author{S. Sarkar$^{1}$, T. Maitra$^{2}$,  Roser Valent{\'\i}$^{3}$,   T. Saha-Dasgupta$^{1}$}
\affiliation{$^{1}$ S. N. Bose National Centre for Basic Sciences, Kolkata, India\\
$^{2}$ Department of Physics, Indian Institute of Technology, Roorkee, India\\
$^{3}$Institut f{\"u}r Theoretische Physik,  Goethe Universit{\"a}t, Frankfurt, Germany}

\date{\today}

\begin{abstract}
Using first-principles density functional calculations, we perform
 a comparative study of two Fe-based spinel compounds, \fecr\ and \Sc. Though
 both systems contain an orbitally active A site with an Fe$^{2+}$ ion,
 their properties are rather dissimilar. Our study unravels the microscopic origin of their behavior driven by the differences in hybridization of Fe $d$ states with Cr/Sc $d$ states and S $p$ states in the two cases. This leads to important differences
in the nature of the magnetic exchanges as well as the nearest versus next nearest neighbor exchange parameter ratios,
resulting into  significant  frustration effects  in \Sc\ which are absent in \fecr.

\end{abstract}
\maketitle
\noindent

Spinel compounds have attracted a lot of attention in the last years due to the intricate interplay 
of spin, charge and orbital degrees of freedom together with intrinsic frustration effects driven by 
their peculiar geometry. A large amount of work has been done on normal spinels of general  formula 
AB$_{2}$X$_{4}$ with tetrahedral AX$_4$ and octahedral BX$_6$ units, and with orbitally 
active B sites like ZnV$_{2}$O$_{4}$\cite{zn,tcher_04,matteo_05,znv},
MnV$_{2}$O$_{4}$ \cite{mn,mnv,perkins_10}, CdV$_{2}$O$_{4}$\cite{zn},
CuIr$_{2}$S$_{4}$\cite{cu} or MgTi$_{2}$O$_{4}$\cite{mg}. Examples of compounds with 
orbitally active A sites also exist, as is
the case of FeCr$_{2}$S$_{4}$ (FCS) and FeSc$_{2}$S$_{4}$ (FSS).
The  Fe$^{2+}$ ion in these cases is in a 3\textit{d}$^{6}$ configuration,
 with a local S=2 moment and a two-fold orbital degeneracy associated with
one hole in a doubly degenerate \textit{e} state of the tetrahedrally
 crystal split \textit{d} levels.
In  FCS, the B cation is magnetic (Cr$^{3+}$ has a spin S=$3/2$)
 while for FSS, the B cation is non-magnetic (Sc$^{3+}$ has a
 filled shell [Ar] configuration). FCS orders magnetically in a ferrimagnetic
 spin arrangement between Fe and Cr moments with a transition
 temperature\cite{min} of $167$K, while FSS does not order magnetically
 down to a measured temperature of 50 mK \cite{krimmel}.
 FCS  shows long range orbital order in polycrystalline samples
 while a glassy freezing has been observed in single crystals.
  FSS, in contrast, has been reported  as an orbital liquid \cite{buttgen}.

Considering the measured Curie-Weiss temperature
 $\left(\Theta_{CW}\right) $ of	 -$200$K (FCS \cite{buttgen})
 and -$45$K (FSS \cite{buttgen}), the frustration parameter defined
 as $f=\frac{-\Theta_{CW}}{T_{N}}$, T$_N$ being the magnetic transition temperature, 
is 1.2 for FCS and larger than 1000 for FSS.
  To our knowledge, the microscopic understanding of this qualitatively
different behavior has not been attempted so far, though 
experimental\cite{min,krimmel,buttgen} as well as  related theoretical work based 
on model Hamiltonians\cite{gangprb,bergman} has been performed.
 One may note that the B sublattice, which forms a pyrochlore lattice of
 corner sharing tetrahedra, is geometrically frustrated in terms of
 nearest neighbor (NN) interactions while the A sublattice forms a diamond lattice
 consisting of two interpenetrating face centered cubic (FCC) sublattices
 which is not frustrated if only NN interactions are assumed.
 In the following, we will investigate the microscopic origin
 of the different behavior between  FSS and FCS in the framework
of  density functional theory (DFT) calculations.
 We considered three different basis sets, namely:  the linear augmented plane wave (LAPW)
method	as implemented in the WIEN2K \cite{wien} code, the muffin-tin orbital(MTO) based
N-th order MTO (NMTO) method\cite{nmto} as implemented in the Stuttgart code and the plane-wave basis
as implemented in the Vienna Ab-initio Simulation Package (VASP)\cite{vasp}. The reliability of the
calculation in the three  basis sets has been cross-checked.

\paragraph*{Crystal Structure -}
Both FCS and FSS crystallize in the
cubic Fd\={3}m structure. The lattice parameters
 of FCS and FSS are reported
 to be $9.99$\AA\ and $10.50$\AA\ \cite{min,crys} respectively,
 showing a $5\%$ expansion in  FSS  due to the
presence of larger Sc$^{3+}$ ions (size	 $\sim$ 0.75 \AA)
 compared to Cr$^{3+}$ ions (size  $\sim$ 0.62 \AA).
 The internal parameter associated with S shows deviations from its ideal
 value of $\frac{1}{4}$, with 0.259 for FCS and 0.255 for FSS \cite{shir,crys}.
 This leads to a trigonal distortion in the BS$_{6}$ octahedra
 measured in terms of the deviation of the S-B-S bond-angle
   from the ideal
90$^{\circ}$ angle; 4.35$^{\circ}$ (FCS) and 2.5$^{\circ}$ (FSS).
 The tetrahedra remain undistorted in both  compounds.

\paragraph*{Electronic Structure -}

\begin{figure}[ht]
\centering
\includegraphics[scale=0.75, draft=false]{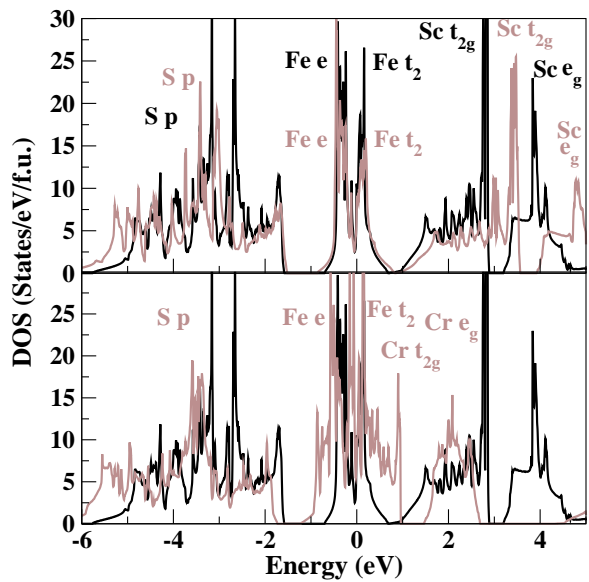}
\caption{(Color online)Non-spin polarized total DOS calculated within GGA, with
the zero of energy set at the fermi level E$_f$. 
Top panel:  DOS of FSS calculated
in the actual crystal structure (dark solid line) and in the crystal structure of FCS (light solid line).
Bottom panel: Comparison DOS of FSS (dark solid line) and FCS (light solid line). The various orbital
contributions are marked for each DOS plots.}
\label{dos}
\end{figure}

\begin{figure}[ht]
\centering
\includegraphics[scale=0.5, draft=false]{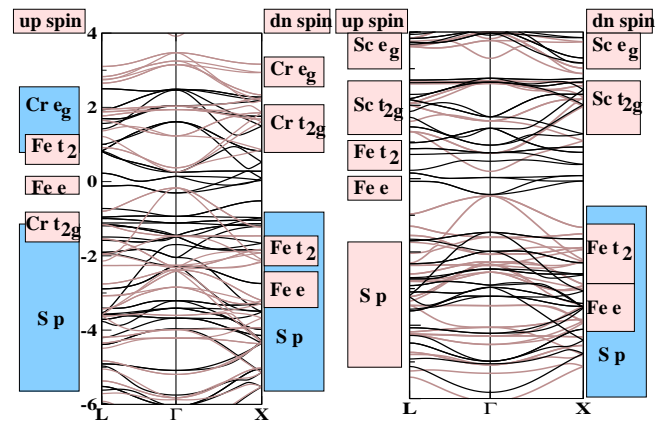}
\caption{(Color online) Spin polarized band structure calculated within GGA.
The zero of  energy is set at E$_f$. Left panel shows band structure of FCS and right panel shows the same for FSS. The dark lines (black in color) represent the bands corresponding to up spin channel and light line (grey in color) represent the bands corresponding to down spin channel. }
\label{band}
\end{figure}

\begin{figure}[ht]
\centering
\includegraphics[scale=0.5, draft=false]{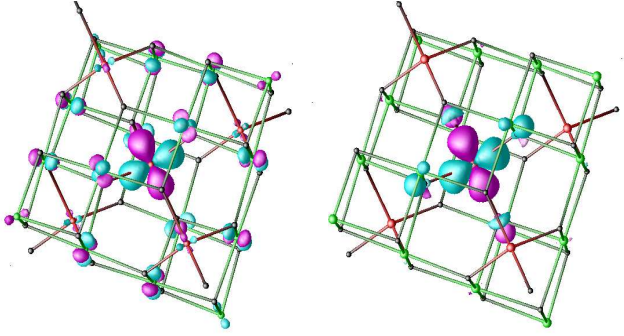}
\caption{Wannier function plot of Fe \textit{d$_{xy}$} orbital for FCS (left panel) and FSS (right panel). Plotted are the constant value surfaces. Two opposite lobes of the wavefunctions are colored differently.}
\label{wannier}
\end{figure}

Fig.\ref{dos} shows  non spin polarized density of states (DOS)
 calculated in the LAPW basis
 with the generalized gradient approximation (GGA) \cite{gga}.
  In order to check the influence on the electronic properties
 of the crystal structure differences between FCS and FSS,
we have also performed calculations for	  FSS assuming the crystal
structure of FCS.
 The top panel of Fig. \ref{dos} shows the DOS of FSS obtained considering
  the actual crystal structure in comparison with the DOS obtained assuming
 the crystal structure of FCS. The bottom panel shows the comparison of DOS
between	 FSS and FCS both in their actual crystal structure.
 We notice that while the change of crystal structure has some effect (a)
in terms of  narrowing	the Fe \textit{d} dominated states at
 the Fermi level (E$_f$) in the actual FSS lattice compared to the results
with the hypothetical lattice and (b)  in the positioning of the empty
 Sc levels spanning the energy window of about 1 eV to 5 eV (Fig. \ref{dos} top
panel),
 the major changes happen upon replacing Sc by Cr (Fig. \ref{dos} bottom
panel). The bandwidth of the Fe-\textit{d} dominated states crossing
E$_f$ is substantially increased  and also there
 is a significant change in the unoccupied region of the spectrum.
 The difference between the electronic
structure of FCS and FSS becomes more evident in the spin polarized
bandstructure shown in Fig. \ref{band}. Although FSS  doesn't
 spin order, such calculations are useful in
understanding the relative positions of Fe and the cation B (Cr or Sc)
 energy levels taking into account the spin degrees of freedom.
 Fe and Cr/Sc \textit{d} states are crystal split, in \textit{e} and
 \textit{t$_{2}$} and \textit{t$_{2g}$} and \textit{e$_{g}$} respectively,
as well as spin split. In the down spin channel,
 the Fe \textit{d} dominated states are completely occupied
 while in the up spin channel Fe \textit{e} states are
partially empty in agreement with the Fe$^{2+}$ nominal valence.
 For Sc, the \textit{d} states are empty in both spin channels with
little shift in the energy scale between the two spin channels,
 proving the essentially non-magnetic character of Sc$^{3+}$.
  The Cr  \textit{d} states are empty in the down spin channel
 and partially occupied in the up spin channel with
   \textit{t$_{2g}$} up spin states
 occupied and	\textit{e$_{g}$} up spin states	 empty
 with a spin splitting of about 2 eV. This is
 in agreement with a ferrimagnetic spin ordering between Fe and Cr.
 The difference between FCS and FSS arises from
 the relative energy positions of Cr and Sc with respect to that of Fe.
While the Sc \textit{d} levels all appear above the Fe \textit{d} states
 with little mixing between them, there exists a rather strong mixing between
Fe \textit{d} and Cr \textit{d} states in the up spin channel.
 It is this Fe-Cr mixing that causes the substantial increase in
 the width of the Fe \textit{d} dominated states crossing E$_f$
 in Fig.~\ref{dos}.  The energy levels of non spin-split Fe and Cr \textit{d} states are found to be within an energy window 0.5 eV, causing near degeneracy between the levels, while the Fe and Sc levels are found to be energetically separated by about 2 eV or more.
\paragraph*{Effective Fe-Fe interaction - }
In order to extract the effective Fe-Fe interactions
 we performed NMTO downfolding calculations.  Starting from a full DFT calculation, the method 
constructs the low energy Hamiltonian defined in an
effective Wannier function basis by integrating out  degrees of freedom that are not of interest ($downfolding$).
 In our downfolding calculations, we have kept active Fe \textit{d} states
 and have downfolded all the other states involving Cr/Sc and S.
Fig.~\ref{wannier} shows the  Fe $d_{xy}$
 Wannier function for FCS and FSS.
 The central region of the Wannier function is shaped according
 to the Fe $d_{xy}$ symmetry while the tails are shaped according
 to the integrated out orbital degrees of freedom {\it e.g.} Cr/Sc
 and S orbitals.  We first  notice  that the Wannier function for
 FCS is much more delocalized compared to that of FSS with
 significant weights at the Cr sites surrounding the central Fe site.
In contrast,  the Wannier function for FSS  is localized with little weight
 on Sc sites and only some weight on the neighboring S sites.

\begin{table}[h]
\caption{Hopping matrix elements (in meV) of FSS and FCS (first two values of each column respectively) and the magnitude of their differences (third value of each column) for the NN ($\left[ \frac{1}{4}\frac{1}{4}\frac{1}{4}\right]$) and
2NN ($\left[ 0\frac{1}{2}\frac{1}{2} \right] $,$\left[ \frac{1}{2}0\frac{1}{2}\right]  $, $\left[ \frac{1}{2}\frac{1}{2}0\right] $).
The matrix elements are listed for distinct entries only. 1,2,3,4 and 5 represent the five \textit{d} orbitals, \textit{d$_{xy}$}, \textit{d$_{yz}$}, \textit{d$_{3z^{2}-1}$}, \textit{d$_{xz}$} and \textit{d$_{x^{2}-y^{2}}$} respectively.}
\centering
\begin{tabular}{|c| c| c| c| c| }
\hline
 m,m$^{\prime}$ & $\left[ \frac{1}{4}\frac{1}{4}\frac{1}{4}\right] $ &	$\left[ 0\frac{1}{2}\frac{1}{2} \right] $ & $\left[ \frac{1}{2}0\frac{1}{2}\right]  $ & $\left[ \frac{1}{2}\frac{1}{2}0\right] $ \\[0.1 cm]
\hline
1,1 & -3 \enskip 60  \bf{63} & \enskip 43 \enskip 12 \bf{31} & \enskip 43 \enskip 12 \bf{31} & -16 -13 \bf{3} \\[0.1 cm]
2,2 & -3 \enskip 60 \bf{63} & -16 -13 \bf{3} & \enskip 43 \enskip 12 \bf{31} & \enskip 43 \enskip 12 \bf{31} \\[0.1 cm]
3,3 & \enskip 11 \enskip 10 \bf{1}  & -6 -1 \bf{5} & -6 -1 \bf{5} & -21 \enskip 2 \bf{23}\\[0.1 cm]
4,4 & -3 \enskip 60 \bf{63} & \enskip 43 \enskip 12 \bf{31}& -16 -13 \bf{3} & \enskip 43 \enskip 12 \bf{31}\\[0.1cm]
5,5 & \enskip 11 \enskip 10 \bf{1} & -16 \enskip 1 \bf{17} & -16 \enskip 1 \bf{17} & -1 -2 \bf{1} \\[0.1 cm]
1,2 & -10 -9 \bf{1} & -22 \enskip 11 \bf{33} & \enskip 46 \enskip 17 \bf{29} & \enskip 22 -11 \bf{33} \\[0.1cm]
1,3& -22 -18 \bf{4} & \enskip 16 \enskip 8 \bf{8} & \enskip 16 \enskip 8 \bf{8}& -11 -22 \bf{11} \\[0.1 cm]
1,4 & -10 -9 \bf{1}& \enskip 46 \enskip 17 \bf{29} & -22 \enskip 11 \bf{33} & \enskip 22 -11 \bf{33} \\[0.1 cm]
1,5 & \enskip \enskip 0 \enskip \enskip 0 \enskip \bf{0}& -18 \enskip 3 \bf{21} & \enskip 18 -3 \bf{21} & 0 0 \bf{0} \\[0.1 cm]
2,3 & 11 9 \bf{2}& 4 11 \bf{7} & 7 -7 \bf{0} & -24 -1 \bf{23}\\[0.1 cm]
2,4 & -10 -9 \bf{1} & \enskip 22 -11 \bf{33} & -22 \enskip 11 \bf{33}& \enskip 46 \enskip 17 \bf{29}\\[0.1 cm]
2,5 & -19 -16 \bf{3} & -7 -19 \bf{12}& \enskip 23 \enskip 5 \bf{18} & \enskip 5 \enskip 8 \bf{3}\\[0.1 cm]
3,4 & \enskip 11 \enskip 9 \bf{2} & -7 7 \bf{14} & 4 11 \bf{7} & 24 1 \bf{23} \\[0.1 cm]
3,5 & \enskip 0 \enskip 0 \enskip \bf{0}& 9 -2 \bf{11}& -9 2 \bf{11} & \enskip 0 \enskip 0 \enskip \bf{0}\\[0.1 cm]
4,5 & \enskip 19 \enskip 16 \bf{3}& -23 -5 \bf{18} & \enskip 7 \enskip 19 \bf{12}& -5 -8 \bf{3}\\[0.1 cm]
\hline
\end{tabular}
\label{hopping}
\end{table}

\begin{figure}[h]
\centering
\includegraphics[scale=0.5, draft=false]{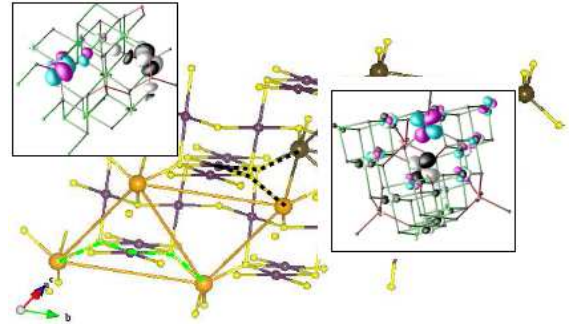}
\caption{(Color online) The 1NN and 2NN interaction path between
 Fe atoms. Small dark and light balls represent B (Sc/Cr)
 and S atoms respectively. Big dark and light balls represent
 Fe atoms belonging to two FCC sublattices constituting the diamond lattice.
 The dashed line in dark and light, represent the 1NN and 2NN paths
 respectively. The inset in the upper-left (lower-right) corner shows
the overlap of the Wannier functions of Fe \textit{d$_{xy}$} placed at
 two Fe atoms in FSS (FCS) separated by 2NN (1NN) distance.}
\label{path}
\end{figure}

The real space Hamiltonian constructed in the effective Wannier
 function basis of Fe is tabulated in Table \ref{hopping}
 considering up to second nearest neighbor (2NN) interactions.
  Focusing on the hopping parameters  listed in Table \ref{hopping} and
their difference (shown in boldface), we find the changes to be most
 significant within the \textit{t$_{2}$} $\left( d_{xy}, d_{yz}, d_{xz} \right)$ block
of the Hamiltonian.
 We observe that
 while for FCS, the Fe-Fe NN hopping integrals
 are larger than the 2NN hopping terms
 (the largest 2NN is about three times smaller than the largest 1NN
 hopping term), the reverse is the case for FSS where
 the 2NN hoppings are larger than the 1NN hoppings (the largest 2NN
 hopping is  twice as big as the 1NN hopping). The 1NN
and 2NN paths between two A ions in a spinel lattice,
 as shown in Fig. \ref{path}, are A-X-B-X-A exchange paths. The
1NN  interaction connects A ions of two FCC sublattices while  the 2NN
 interaction connects  A ions within the same FCC sublattice.
 The large value of the 2NN interaction
  can therefore generate strong frustration.

 The 1NN hopping path as marked
 in Fig.~\ref{path},
includes Fe-B-Fe, Fe-S-Fe and S-B-S bond angles of about 60$^{\circ}$, 80$^{\circ}$	and 90$^{\circ}$ respectively, while
the corresponding bond angles for the 2NN hopping
 paths are found to be close to 120$^{\circ}$, 130$^{\circ}$
and 90$^{\circ}$ respectively\cite{footnote}.
For the 1NN it is therefore the direct Fe-B hybridization
 that becomes important,
 with anions playing little role
 while for the
2NN interaction, the anion mediated (Fe-S-Fe) exchange becomes important.
The fact that the 1NN interaction is
  strong in FCS and the 2NN interaction is strong in FSS
  is supported by the plot of the
 Wannier functions for two 1NN Fe sites (top left panel of Fig.~\ref{path})
and two 2NN Fe sites (bottom right panel of Fig.~\ref{path}).
  For FCS, we find a clear overlap of Cr-like tails between two
 Wannier functions, while for FSS  the S-like
tails  point to each other.

The exchange interaction may be derived from the hopping integrals
through the use of a superexchange like formula. This however needs the knowledge
of the appropriate charge transfer energy, which is difficult to estimate
because of complicated hopping
paths.	We therefore preferred to compute the effective
magnetic exchange interactions between Fe ions in terms of total energy calculations
of different spin arrangements of Fe and mapping the total energies to an Ising like model
defined in terms of Fe spins. For this purpose, spin-polarized calculations 
were carried out with a plane wave basis as implemented in VASP and
with the choice of the GGA exchange-correlation functional.
While admittedly such calculations are
faced with several difficulties like the choice	 of spin configurations
in supercells, particularly since it involves small energies, it is expected to 
provide us with relative strength of various
exchange interactions as well as some order of magnitude estimates.
For FSS, our calculations gave  J$_{1}$ = -0.01 meV (1NN) and
J$_{2}$ = -0.37 meV (2NN)  with J$_{2}$/J$_{1}$ = 37;
 the 2NN interaction dominates
the NN interaction, as already inferred from the hopping parameters.
This is in agreement with the findings of  neutron scattering
measurements \cite{krimmel}.  For FCS we obtained J$_{1}$ = 6 meV and J$_{2}$ = 2.5 meV
 both being of ferromagnetic nature,  in
agreement with the observed ferromagnetic spin ordering within the Fe sublattice.
The NN interaction dominates over the 2NN neighbor interaction in this case, with J$_{2}$/J$_{1}$ = 0.4
in sharp contrast with that of FSS.

\paragraph*{Spin-Orbit Coupling -}

\begin{table}[ht]
\caption{Magnetic moments of  Fe  and B(Cr/Sc) ions in $\mu_{B}$ and anisotropy energy in meV/Fe .}
\centering
\begin{tabular}{c cc cc c}
\hline \hline
 & \multicolumn{2}{c}{Fe} & \multicolumn{2}{c}{B(Cr/Sc)} & Anisotropy \\
 & Orbital & Spin & Orbital & Spin & energy \\
 & moment & moment & moment & moment & meV/Fe \\[0.5 ex]
\hline
FCS &  -0.13 &-3.27  & -0.03 & 2.69 & 10\\
FSS & -0.14 & -3.44& 0.0 & 0.05 & 6\\ [1ex]
\hline\hline
\label{moment}
\end{tabular}
\end{table}
Due to the presence of unquenched orbital degrees of freedom on the
 Fe sites, the importance of the spin-orbit (SO) coupling in these
 compounds has been discussed\cite{gangprb} in the past.
  An important quantity in this context is the relative strength
of the SO coupling parameter, $\lambda$, with respect to the
dominant spin exchange.
In Table \ref{moment} we show the magnetic moments at the Fe
 and B (Cr/Sc) site obtained from a  GGA+U+SO calculation in LAPW basis 
carried out for FCS and FSS  by considering  a J = 1eV (Hund's coupling) and
U = 2.5 eV at the
 Fe site due to the Coulomb renormalization of the spin-orbit splitting,
 as found previously \cite{fecr2s4}.
A rather large moment of 0.13 - 0.14 $\mu_{B}$ pointing
 along the same direction as the spin moment has been obtained
 at the Fe site for both FCS and FSS.  Such  values are surprisingly large
given the fact that the orbitally active levels of Fe are
 \textit{e} levels. This has been rationalized in terms of finite coupling between
 Fe \textit{e} and empty \textit{t$_{2}$} orbitals~\cite{fecr2s4}.
 Table \ref{moment} also lists the magneto-crystalline
anisotropy energy obtained as the total energy difference
 between the calculations with the spin
 quantization along [001] and [110]. The anisotropy energy is
 found to be almost 2 times larger for	FCS compared to FSS,
indicating stronger spin-orbit interaction in FCS.
 The strength of the spin-orbit interaction depends
on the energy level separation ($\Delta$) between  Fe \textit{e} and \textit{t$_{2}$}.
 We have estimated $\Delta$ from the NMTO- downfolding calculations in the effective
 Fe only basis, and obtain\cite{footnote2} $\Delta=0.46$ eV for
 FSS and $\Delta= 0.20$ eV for
FCS.
 Using second order perturbation theory\cite{vallin} as
 considered in Ref. \cite{gangprb}, the spin-orbit coupling parameter
 is given by $\lambda \sim \frac{6 \lambda^{2}_{0}}{\Delta}$,
 where $\lambda_{0}$ is the atomic spin-orbit coupling constant, estimated
 to be 0.01 eV \cite{testelin}. We obtain  $\lambda$ = 1.3 meV (FSS)
 and 3 meV (FCS). Considering the dominant magnetic interaction into account,
  $J\over \lambda$ is $\gg$ 1 in  FCS and $\ll$ 1 in  FSS. As discussed in Refs.[\onlinecite{gangprb}], these two
situations will give rise to very different ground states, an magnetically
ordered state for  $J\over \lambda$ $\gg$ 1 and a spin orbital singlet for $J\over \lambda$ $\ll$ 1.

To conclude, we have carried out DFT calculations to provide
 a microscopic understanding of the dissimilar
 behavior of spinel compounds FCS and FSS,
 both having orbitally active A ions.
We found that this originates from the
 difference in the hybridization between Fe \textit{d} states and 
 B (B=Cr/Sc) states and S $p$ states. This not only affects the magnitude 
of magnetic exchanges, but also the relative importance of different magnetic 
exchanges: A contrasting value of J2/J1 of 37 in the case of the Sc compound to a value of 0.4 in the  case of the Cr compound. Moreover, the J's are antiferromagnetic for the Sc systems and ferromagnetic for the Cr system. This leads to important frustration effects in the Sc compound which are absent in the Cr compound. In our entire analysis, we have not considered the effect
 of Jahn-Teller (JT) interactions. 
 Though crystallographically no signature for static JT order has been found,
 there could  be dynamic JT effects. This will be taken up in a future study.

{\it Acknowledgements.-} TSD would like acknowledge support from DST through AMRU project.
SS thanks CSIR for financial support. RV thanks the DFG for financial support through
the SFB/TRR 49 program.

\end{document}